\documentclass[12pt,fleqn]{article}

\usepackage{epsfig}
\usepackage{amsmath}
\usepackage{amsfonts}
\usepackage{times}
\usepackage{natbib}

\makeatletter
\renewcommand\@biblabel[1]{#1.}
\makeatother

\title{\sffamily\textbf{\vspace*{-20mm}\\
Symmetric competition as a general model for single-species adaptive dynamics
}}
\author{Michael Doebeli$^1$ \& Iaroslav Ispolatov$^{1,2}$\\\\
\vspace{-3mm}\normalsize $^1$Department of Zoology and Department of Mathematics\\
\vspace{-2mm}\normalsize University of British Columbia, 7280 University Boulevard\\
\vspace{-2mm}\normalsize Vancouver B.C. Canada, V6T 1Z4\\
\vspace{-2mm}\normalsize $^2$Departamento de F\'isica\\
\vspace{-2mm}\normalsize Universidad de Santiago de Chile\\
\vspace{10mm}\normalsize Casilla 307, Santiago 2, Chile\\
\vspace{-2mm}\normalsize $^{1,2}$ To whom correspondence may be addressed. \\
\vspace{-2mm}\normalsize E-mail: doebeli@zoology.ubc.ca, jaros007@gmail.com.
}
\vspace{10mm}
\date{\vspace{10mm}\normalsize\today}

\bibpunct{(}{)}{,}{"a"}{}{;}

\newtheorem{theorem}{Theorem}[section]

\newtheorem{proposition}[theorem]{Proposition}
\newtheorem{corollary}[theorem]{Corollary}

\newenvironment{proof}[1][Proof]{\begin{trivlist}
\item[\hskip \labelsep {\bfseries #1}]}{\end{trivlist}}

\newcommand{\qed}{\nobreak \ifvmode \relax \else
      \ifdim\lastskip<1.5em \hskip-\lastskip
      \hskip1.5em plus0em minus0.5em \fi \nobreak
      \vrule height0.75em width0.5em depth0.25em\fi}
      
\begin{document}
\maketitle

\newpage
{\bf \large Abstract}
\vskip 0.5cm

{ Adaptive dynamics is a widely used framework for modeling long-term evolution of continuous phenotypes. It is based on invasion fitness functions, which determine selection gradients and the canonical equation of adaptive dynamics. Even though the derivation of the adaptive dynamics from a given invasion fitness function is general and model-independent, the derivation of the invasion fitness function itself requires specification of an underlying ecological model. Therefore, evolutionary insights gained from adaptive dynamics models are generally model-dependent. Logistic models for symmetric, frequency-dependent competition are widely used in this context. Such models have the property that the selection gradients derived from them are gradients of scalar functions, which reflects a certain gradient property of the corresponding invasion fitness function. We show that any adaptive dynamics model that is based on an invasion fitness functions with this gradient property can be transformed into a generalized symmetric competition model. This provides a precise delineation of the generality of results derived from competition models. Roughly speaking, to understand the adaptive dynamics of the class of models satisfying a certain gradient condition, one only needs a complete understanding of the adaptive dynamics of symmetric, frequency-dependent competition. We show how this result can be applied to number of basic issues in evolutionary theory.}
 
\newpage

\section{Introduction}
\vskip 0.5cm

Adaptive dynamics (\cite{metz_etal1996, geritz_etal1998, dieckmann_law1996}) has emerged as a widely used framework for modeling long-term evolution of continuous phenotypes. The basic ingredient of an adaptive dynamics model is the invasion fitness function (\cite{metz_etal1992}), which describes the ecological growth rate of rare mutant phenotypes in a given resident community, which is assumed to persist on a community-dynamical attractor. The invasion fitness function determines the selection gradients, which are in turn the core ingredient for deriving the canonical equation (\cite{dieckmann_law1996}) for the adaptive dynamics of the phenotypes under consideration. Following this basic recipe, adaptive dynamics models have been constructed for a plethora of different ecological settings, and have been used to analyze a number of interesting and fundamental evolutionary scenarios, such as evolutionary cycling in predator-prey arms races (\cite{dieckmann_etal1995, marrow_etal1996}), evolutionary diversification (\cite{dieckmann_doebeli1999, dieckmann_etal2004, doebeli2011}) and evolutionary suicide {\cite{gyllenberg_parvinen2001, parvinen2005}}. Even though the derivation of the adaptive dynamics from a given invasion fitness function is general and model-independent, the derivation of the invasion fitness function itself requires specification of an underlying ecological model. Therefore, evolutionary insights gained from adaptive dynamics models are generally tied to a specific ecological setting, and hence model-dependent. 

One particular ecological model that has been often used to derive invasion fitness functions and adaptive dynamics is the symmetric logistic competition model, which in fact is the most popular ecological model in the theory of ecology and evolution. The basic form of this model is
\begin{align}
\frac{dN}{dt}=rN\left(1-\frac{N}{K}\right),
\end{align}
where $N$ is population density, and $r$ and $K$ are parameters describing the intrinsic per capita growth rate and the equilibrium population size, respectively. $K$ is often called the carrying capacity of the population, but it useful to note that $K$ can also be interpreted as a property of individuals, i.e., as a measure of how well individuals cope with competition, as expressed in the per capita death rate $rN/K$. The ecological model (1) can be used to construct a well-known adaptive dynamics model by assuming that the carrying capacity $K(x)$ is a positive function that depends on a continuous, 1-dimensional phenotype $x$ (e.g., body size), and that the competitive impact between individuals of phenotypes $x$ and $y$ is given by a competition kernel $\alpha(x,y)$. For simplicity, it is often assumed that the intrinsic growth rate $r$ is independent of the phenotype $x$, and hence is set to $r=1$. To derive the corresponding invasion fitness function, it is assumed that a resident type $x$ is at its ecological equilibrium density $K(x)$. The dynamics of the density $N(y)$ of a mutant type $y$ is then given by 
\begin{align}
\frac{dN(y)}{dt}=N(y)\left(1-\frac{\alpha(x,y)K(x)+\alpha(y,y) N(y)}{K(y)}\right),
\end{align}
where $\alpha(x,y)K(x)$ is the competitive impact that the resident population exerts on mutant individuals. Assuming that the mutant is rare, $N(y)\approx 0$, the per capita growth rate of the mutant type $y$ in the resident $x$ is
\begin{align}
f(x,y)=1-\frac{\alpha(x,y)K(x)}{K(y)}.
\end{align}
This is the invasion fitness function for the given ecological scenario (\cite{doebeli2011}). The corresponding selection gradient is
\begin{align}
s(x)=\left.\frac{\partial f(x,y)}{\partial y}\right\vert_{y=x}.
\end{align}
The selection gradient in turn is the main determinant of the adaptive dynamics. More precisely, for 1-dimensional traits $x$ the adaptive dynamics is 
\begin{align}
\frac{dx}{dt}=m(x)\cdot s(x),
\end{align}
where $m(x)>0$ is a scalar quantity that reflects the rate at which mutations occur (\cite{metz_etal1996, geritz_etal1998, dieckmann_law1996}). 

In particular, singular points, i.e., equilibrium points of the adaptive dynamics, are solutions $x^*$ of $s(x^*)=0$. Symmetric competition models are characterized by the assumption that $\partial \alpha(x,y)/\partial y\vert_{y=x}=0$ for all $x$, it is also usually assumed that $\alpha(x,x)=1$.
 In this case, $s(x)=K'(x)/K(x)$, and singular points $x^*$ are given as solutions of $K'(x^*)=0$, and hence are given as local maxima and minima of the carrying capacity. Moreover, singular points that are local maxima are attractors of the adaptive dynamics, i.e., $ds/dx(x^*)<0$. For logistic models it is often assumed that $K(x)$ is unimodal, attaining a unique maximum at some trait value $x_0$. $K(x)$ then represents a (global) stabilizing component of selection for $x_0$, and the adaptive dynamics (5) converges to $x_0$. 

In contrast, the competition kernel $\alpha(x,y)$ generally represents the frequency-dependent component of selection. For symmetric competition it is usually assumed that the effect of competition decreases with increasing phenotypic distance $\vert x-y\vert$, which implies negative frequency dependence, because it confers a competitive advantage to rare phenotypes. However, the opposite is also possible, so that $\alpha(x,y)$ has a local minimum as a function of $y$ at $y=x$, which can occur for example in models with explicit resource dynamics (\cite{ackermann_doebeli2004}). While the carrying capacity is the sole determinant of the singular points and their convergence stability for the adaptive dynamics (5), the competition kernel comes into play when determining the evolutionary stability of the singular point $x^*$.
Evolutionary stability of $x^*$ is determined by  the second derivative of the invasion fitness function at the singular point:
\begin{align}
\left.\frac{\partial^2 f(x^*,y)}{\partial y^2}\right\vert_{y=x^*}=\frac{K''(x^*)}{K(x^*)}-\left.\frac{\partial^2 \alpha(x^*,y)}{\partial y^2}\right\vert_{y=x^*}.
\end{align}
The singular point $x^*$ is evolutionarily stable if and only if expression (6) is negative, and it is clear that if $\partial^2 \alpha(x^*,y)/\partial y^2\vert_{y=x^*}$ is negative enough, then this condition will not be satisfied, and instead $x^*$ will be evolutionarily unstable. In particular, the distinction between convergence stability and evolutionary stability makes it clear that it is possible for the singular point $x^*$ to be both convergent stable and evolutionarily unstable. In this case, $x^*$ is called an evolutionary branching point, for such points are potential starting points for evolutionary diversification. The phenomenon of evolutionary branching is an iconic feature of adaptive dynamics and has been studied extensively (\cite{geritz_etal1998, doebeli2011}).

It is straightforward to extend the symmetric logistic competition model to multi-dimensional phenotype spaces (\cite{dieckmann_law1996,leimar2009,doebeli_ispolatov2010,doebeli2011}). In this case, $x\in\mathbb{R}^{m}$ is a $m$-dimensional vector, where $m$ is the dimension of phenotype space, and $K(x):\mathbb{R}^{m}\rightarrow \mathbb{R}$ is a scalar function, as is the competition kernel $\alpha(x,y):\mathbb{R}^{m}\times\mathbb{R}^{m}\rightarrow\mathbb{R}$. For symmetric competition, it is assumed that the partial derivatives $\partial \alpha(x,y)/\partial y_i\vert_{y=x}=0$ for all $x$ and all $i=1,...,m$ (as well as, without loss of generality, $\alpha(x,x)=1$ for all $x$). The corresponding invasion fitness function $f(x,y):\mathbb{R}^{m}\times\mathbb{R}^{m}\rightarrow\mathbb{R}$ again has the form (3), and the selection gradient is given as a vector-valued function $s(x)=(s_{1}(x),\ldots,s_{m}(x)),:\mathbb{R}^{m}\rightarrow\mathbb{R}$, where
\begin{align}
s_i(x)=\left.\frac{\partial f(x,y)}{\partial y_i}\right|_{y=x},\quad i=1,...,m
\end{align}
The adaptive dynamics is then given by a system of $m$ coupled differential equations
\begin{align}
\frac{dx}{dt}=M(x)\cdot s(x),
\end{align}
where $M(x)$ is the $m\times m$ mutational variance-covariance matrix, specifying the rate and magnitude at which mutations occur in the various trait components $x_i$, as well how mutations in different trait components are correlated (\cite{leimar2009, doebeli2011}). The matrix $M(x)$ is typically assumed to be symmetric and positive definite.

For symmetric competition models, it is easy to see that the $m$ components of the selection gradient are
\begin{align}
s_i(x)=\left.\frac{\partial f(x,y)}{\partial y_i}\right|_{y=x}=\frac{1}{K(x)}\cdot\frac{\partial K(x)}{\partial x_i},
\end{align}
for $i=1,...,m$. Thus, the selection gradient is the gradient of a scalar function,
\begin{align}
s(x)=\nabla S(x),
\end{align}
with $S(x)=\ln(K(x))$. 
Thus, just as in the 1-dimensional symmetric competition model, the selection gradients in multi-dimensional generalizations of symmetric competition are essentially gradients of the carrying capacity function $K(x)$ (i.e., gradients of the stabilizing component of selection). The selection gradients therefore induce an evolutionary hill-climbing process towards local maxima of the carrying capacity. The adaptive dynamics (8), resulting from applying the mutational variance-covariance matrix to the selection gradient, is a ``warped'' version of the hill-climbing process generated by the selection gradients. If the mutational matrix $M(x)$ is positive definite, this warped hill-climbing process is essentially equivalent to the unwarped version defined by the selection gradients alone. In particular, in this case the adaptive dynamics (8) also converges to local maxima of $K(x)$.

It is worth noting that the structure of the canonical equation (8) is similar to other general equations for evolutionary dynamics, such as those introduced by \cite{lande1979} and those introduced by \cite{nowak_sigmund1990}. But the assumptions underlying the canonical equation (8), and in particular the notion of invasion fitness, are unique features of adaptive dynamics. In general, invasion fitness functions $f(x,y)$ can be derived for multidimensional phenotypes in each of a number of interacting and coexisting species for many different ecological scenarios ({\cite{dieckmann_law1996,leimar2009,doebeli2011}). In each of the interacting species, invasion fitness is the long-term per capita growth rate of rare mutant types $y$ in a population in equilibrium that is monomorphic for the resident type $x$. Here we consider the following question: under what conditions does an arbitrary invasion fitness function for multi-dimensional phenotypes in a single species has the form of an invasion fitness function derived from a logistic symmetric competition model? This is a relevant question because symmetric competition models have been used for a long time as a basic metaphor to generate ecological and evolutionary insights. It is therefore of interest to understand how universal such models are. For example, in \cite{doebeli_ispolatov2010} it has recently been shown that evolutionary branching, and hence adaptive diversification, becomes more likely in symmetric competition models if the dimension of phenotype space is increased, and it would be useful to know whether this applies to other models. In the present paper  we give a general and precise condition for any given invasion fitness to be equivalent to the invasion fitness function derived from a symmetric competition model.  We also show how this result can be applied to shed light on some general issues in adaptive dynamics theory, such as the notion of frequency-dependent selection, the relationship between symmetric and asymmetric competition, the existence of complicated evolutionary dynamics, and the problem of evolutionary stability in single-species models. 

\vskip 1cm
\section{A condition for universality}
\vskip 0.5cm

We consider an invasion fitness function $f(x,y)$ for a single species, in which the multidimensional trait $x=(x_1,\ldots,x_m)$ denotes the resident trait, and the vector $y=(y_1,\ldots,y_m)$ denotes the mutant trait. Recall that the selection gradient $s(x)$ is a vector 
\begin{align}
s(x)=(s_1(x),\ldots,s_m(x)),
\end{align}
where
\begin{align}
s_i(x)=\left.\frac{\partial f(x,y)}{\partial y_i}\right\vert_{y=x}
\end{align}
\cite{dieckmann_law1996,geritz_etal1998}. Thus, for a given resident trait vector $x$, the selection gradient $s(x)$ is the gradient of the fitness landscape given by $f(x,y)$, but it is important to note that $s(x)$, which is a vector-valued function on an $m$-dimensional space, is defined as the gradient with respect to $y$ of a scalar function defined on a $2m$-dimensional $(x,y)$ space. In particular, the vector field $s(x)$ is in general not the gradient field of a scalar function defined on $m$ variables. 

The latter case is captured in the following Definition: We call the selection gradient $s(x)$ a {\it gradient field} if it can be obtained as the derivative of a scalar function $S(x)$, i.e., if there is a function $S(x)$ such that 
\begin{align}
s(x)=\nabla S(x),
\end{align}
i.e., such that 
\begin{align}
s_i(x)=\frac{\partial S(x)}{\partial x_i}\quad \text{for}\quad i=1,...,m.
\end{align}
The following proposition describes the conditions under which a given invasion fitness function is equivalent to an invasion fitness function derived from a symmetric competition model.
\begin{proposition}
The selection gradient $s(x)$ of an invasion fitness function $f(x,y)$ is a gradient field if and only if the invasion fitness function is of the form 
\begin{align}
f(x,y)=1-\frac{\tilde \alpha(x,y)\tilde K(x)}{\tilde K(y)}
\end{align}
for some scalar functions $\tilde K(x)$ and $\tilde \alpha(x,y)$, such that $\tilde K(x)>0$, $\tilde \alpha(x,x)=1$ for all $x$, and $\partial \tilde \alpha(x,y)/\partial y\vert_{y=x}=0$ for all $x$.
\end{proposition}

\begin{proof}
If the invasion fitness function $f(x,y)$ has the form (15), let $S(x)=\ln \tilde K(x)$. Then it is easy to calculate that
\begin{align}
s(x)=\frac{\nabla\tilde K(x)}{\tilde K(x)}=\nabla S(x),
\end{align}
where $\nabla$ is short for $(\partial/\partial x_1,\ldots,\partial/\partial  x_m)$. Hence the selection gradient is a gradient field. 

Conversely, if the selection gradient is a gradient field, $s(x)=\nabla S(x)$, then let 
\begin{align}
\tilde K(x)=\exp\left[S(x)\right],
\end{align}
and 
\begin{align}
\tilde \alpha(x,y)=\left[1-f(x,y)\right]\frac{\tilde K(y)}{\tilde K(x)}.
\end{align}
Note that $\tilde K(x)>0$ and $\tilde \alpha(x,x)=1$ for all $x$, because $f(x,x)=0$ for any invasion fitness function (i.e., the long-term per capita growth rate of individuals with the resident phenotype must be 0, because the resident, assumed to exist on an equilibrium, neither goes extinct nor increases without
bounds). Moreover, using the fact that $s(x)=\nabla S(x)=
\nabla \tilde K(x)\cdot \tilde K(x)^{-1}$ by construction, it is easy to check that $\partial \tilde \alpha(x,y)/\partial y\vert_{y=x}=0$ for all $x$, and that expression (15) holds for the invasion fitness function $f(x,y)$. \qed
\end{proof}

Thus, any invasion fitness function for which the selection gradient is a gradient field can be viewed as the invasion fitness function of a generalized Lotka-Volterra model for symmetric, frequency-dependent competition. As mentioned in the previous section, in this case the adaptive dynamics is a hill-climbing process on a fixed fitness landscape determined by the (generalized) carrying capacity $\tilde K(x)$ defined by (17), and singular points of the adaptive dynamics are given as local extrema of $\tilde K(x)$. In particular, a singular point of the adaptive dynamics is convergent stable if and only if it is a local maximum of $\tilde K(x)$. 

Next, we derive a general condition for a selection gradients to be a gradient field. For this we recall that a geometric object is called simply connected if any closed path can be shrunk continuously to a single point within that object. Thus, roughly speaking an object is simply connected if it doesn't have any holes. A typical example is Euclidean space $\mathbb{R}^n$ for any $n$.

\begin{proposition}
Assume that the phenotype space (i.e., the set of all attainable, m-dimensional phenotype
vectors $x$) is simply connected. Then the selection gradient $s(x)$ of an invasion fitness function $f(x,y)$ is a gradient field if and only if the invasion fitness function satisfies  
the following {\it gradient condition}: 
\begin{align}
\left.\frac{\partial^2 f(x,y)}{\partial x_i\partial y_j}\right\vert_{y=x}= \left.\frac{\partial^2 f(x,y)}{\partial x_j\partial y_i}\right\vert_{y=x}
\end{align}
for all $i,j$.
\end{proposition}

\begin{proof}
First of all, we note that in any case
\begin{align}
\frac{\partial s_i(x)}{\partial x_j}=\left.\frac{\partial^2 f}{\partial x_j\partial y_i}\right\vert_{y=x}+\left.\frac{\partial^2 f}{\partial y_j\partial y_i}\right\vert_{y=x}.
\end{align}
Therefore,
\begin{align}
D_{ij}(x):=\frac{\partial s_i(x)}{\partial x_j}-\frac{\partial s_j(x)}{\partial x_i}=\left.\frac{\partial^2 f}{\partial x_j\partial y_i}\right\vert_{y=x}-\left.\frac{\partial^2 f}{\partial x_i\partial y_j}\right\vert_{y=x}.
\end{align}
If the selection gradient is a gradient field, $s(x)=\nabla S(x)$, then 
\begin{align}
D_{ij}(x)=\frac{\partial^2S(x)}{\partial x_j\partial x_i}-\frac{\partial^2S(x)}{\partial x_i\partial x_j}=0
\end{align}
for all $x$, as claimed. Conversely, according to basic theory of differential forms (\cite{spivak1999}), the condition $D_{ij}=0$ is equivalent to saying that the differential of the selection gradient is 0, i.e., that the selection gradient is a closed form. A closed form is exact if it is the differential of a function, and exactness of closed forms is captured by the first De Rham cohomology group (\cite{spivak1999}). For simply connected spaces this group is trivial, which implies that every closed form is exact, i.e., every closed form is the differential of a function, and hence a gradient field. \qed
\end{proof}

Together, the two propositions above yield the following
\begin{corollary}
Let $f(x,y)$ be an invasion fitness function defined for a single species on a simply connected phenotype space of arbitrary dimension. Then the function $f(x,y)$ is  equivalent to the invasion fitness derived from a generalized Lotka-Volterra model for symmetric, frequency-dependent competition if and only if   
\begin{align}
\left.\frac{\partial^2 f(x,y)}{\partial x_i\partial y_j}\right\vert_{y=x}= \left.\frac{\partial^2 f(x,y)}{\partial x_j\partial y_i}\right\vert_{y=x}
\end{align}
for all $i,j$. If this condition is satisfied, the generalized carrying capacity and competition kernel are given by expressions (17) and (18).
\end{corollary}

\vskip 1cm
\section{Applications}
\vskip 0.5cm

We illustrate the potential usefulness of the theory presented in the previous section with some examples. 

\vskip 0.5cm
\subsection{Definition of frequency-dependent selection}
The first application concerns the conceptual issue of frequency-dependent selection, a fundamental and much debated topic in evolutionary theory. For adaptive dynamics models, it has been argued by \cite{heino_etal1999}  that selection should be considered frequency-independent (or trivially frequency-dependent) if the resident phenotype $x$ only enters the invasion fitness function through a scalar function describing the resident's effect on the environment (e.g., the resident population size). Otherwise, selection is frequency-dependent. If an adaptive dynamics model is given by an invasion fitness function whose selection gradient is a gradient field, this notion of frequency dependence can be made mathematically precise in the context of the generalized competition model (15): selection is frequency-independent if and only if the generalized competition kernel $\tilde \alpha(x,y)$ given by (18) is a constant (equal to 1), i.e., if and only if 
\begin{align}
f(x,y)=1-\frac{\tilde K(x)}{\tilde K(y)},
\end{align}
where $\tilde K(x)$ is the generalized carrying capacity (17). Because the generalized carrying capacity is entirely defined in terms of the invasion fitness, this leads to a model-independent definition of frequency independence (and hence of frequency dependence).

\vskip 0.5cm
\subsection{Universality of symmetric competition in 1-dimensional phenotype space}
The second application concerns the adaptive dynamics in 1-dimensional phenotype spaces. In this case, any differentiable invasion fitness function trivially satisfies the gradient condition (19), i.e., the selection gradient $s(x)$ is always an integrable function, and hence any 1-dimensional adaptive dynamics model that is based on a differentiable invasion fitness function is equivalent to the adaptive dynamics of symmetric logistic competition. We illustrate this by considering the invasion fitness function for a 1-dimensional trait under asymmetric competition,
\begin{align}
f(x,y)=1-\frac{\alpha(x,y)K(x)}{K(y)}, 
\end{align}
where the competition kernel $\alpha(x,y)$ still has the property that $\alpha(x,x)=1$ for all $x$, but $\partial \alpha(x,y)/\partial y\vert_{y=x}\neq0$ in general. Such models have been considered in the literature (e.g. \cite{taper_case1992,kisdi1999,doebeli_dieckmann2000, doebeli2011}) and reflect the assumption of an  intrinsic advantage of one phenotypic direction, such as an intrinsic advantage to being higher or larger when competition between plant individuals is affected by access to sunlight. As in symmetric competition models, the function $K(x)$ is the carrying capacity function.

In this case, it follows that 
\begin{align}
s(x)=\left.\frac{\partial f(x,y)}{\partial y}\right\vert_{y=x}=-\left.\frac{\partial \alpha(x,y)}{\partial y}\right\vert_{y=x}+\frac{K'(x)}{K(x)}.
\end{align}
Let 
\begin{align}
A(x)=\left.\frac{\partial \alpha(x,y)}{\partial y}\right\vert_{y=x}
\end{align}
and consider the generalized carrying capacity 
\begin{align}
\tilde K(x)=\exp\left[-\int^x A(x')dx'+\ln\left(K(x)\right)\right].
\end{align}
and the generalized competition kernel 
\begin{align}
\tilde \alpha(x,y)=\left(1-f(x,y)\right)\frac{\tilde K(y)}{\tilde K(x)}.
\end{align}
Then $\tilde K(x)>0$ for all $x$, $\tilde \alpha(x,x)=1$ for all $x$, and $\partial\tilde \alpha(x,y)/\partial y\vert_{y=x}=0$ for all $x$. Also
\begin{align}
f(x,y)=1-\frac{\tilde \alpha(x,y)\tilde K(x)}{\tilde K(y)},
\end{align}
and hence the invasion fitness function for asymmetric competition is equivalent to the invasion fitness function of a symmetric competition model, in which the original asymmetry in the competition kernel is shifted onto the generalized carrying capacity function $\tilde K(x)$.

\vskip 0.5cm
\subsection{Cyclic adaptive dynamics in asymmetric single-species models}
Third, to further illustrate the implications of the gradient condition (19), we present an example of the adaptive dynamics that may occur when the gradient condition (19) is not satisfied. For this we consider an asymmetric competition model with a 2-dimensional phenotype space defined by a carrying capacity function $K(x)$ and a competition kernel $\alpha(x,y)$, where $x=(x_1,x_2)$ and $y=(y_1,y_2)$ are 2-dimensional traits, as follows:
\begin{align}
K(x)=\exp\left(-\frac{x_1^4+x_2^4}{2\sigma_K^4}\right)
\end{align}
\begin{align}
\alpha(x,y)=&\exp\left(-\frac{(x_1-y_1)^2+(x_2-y_2)^2}{2\sigma_\alpha^2}\right)\\
\nonumber&\times\exp\left(c_1(x_1y_2-x_2y_1)+c_2(x_1(y_1-x_1)+x_2(y_2-x_2))\right).
\end{align}
We will not try to biologically justify this choice of functions, except to note that the competition kernel (32) has been chosen so that for $i=1,2$, $\partial \alpha(x,y)/\partial {y_i}\vert_{y=x}\neq0$. Biologically, this means that there are epistatic interactions between the trait components $x_1$ and $x_2$, and as we will see, such interactions can be the source of complicated evolutionary dynamics even in a single evolving species. Note that the carrying capacity (31) has a maximum at $(0,0)$, that the competition kernel $\alpha$ given by (32) retains the property that $\alpha(x,x)=1$ for all $x$, and that for $c_1=c_2=0$, the competition kernel becomes symmetric. For $c_1\neq0$ or $c_2\neq0$, and in contrast to the 1-dimensional case for asymmetric competition kernels discussed above, the resulting invasion fitness function $f(x_1,x_2,y_1,y_2)$ does not satisfy the gradient condition (19), and hence is not equivalent to the invasion fitness of a symmetric competition model. This can be seen by directly checking condition (19), or by considering the corresponding adaptive dynamics given by the selection gradients
\begin{align}
s_1(x)=&\left.\frac{\partial f(x,y)}{\partial y_1}\right\vert_{y=x}=\left.-\frac{\partial \alpha(x,y)}{\partial y_1}\right\vert_{y=x}+\frac{\partial K(x)}{\partial x_1}\cdot\frac{1}{K(x)},\\
s_2(x)=&\left.\frac{\partial f(x,y)}{\partial y_2}\right\vert_{y=x}=\left.-\frac{\partial \alpha(x,y)}{\partial y_2}\right\vert_{y=x}+\frac{\partial K(x)}{\partial x_2}\cdot\frac{1}{K(x)}.
\end{align}
It is easy to see that this adaptive dynamical system has a singular point at $x^*=(0,0)$ (note that $\partial \alpha(x,y)/\partial {y_i}\vert_{y=x=x^*}=0$ for $i=1,2$), and that the Jacobian at the singular point $x^*$ has complex eigenvalues $\pm ic_1 -  c_2$. In particular, if $c_1\neq0$, the adaptive dynamics has a cyclic component in the vicinity of the singular point $x^*$, and it follows that if $c_1\neq0$, the selection gradient $s=(s_1,s_2)$ cannot be a gradient field because the gradient condition is not satisfied. In fact, the adaptive dynamics given by the selection gradients (33) and (34) can exhibit a stable limit cycle, as is illustrated in Figure 1. Cyclic evolutionary dynamics are known to occur in the adaptive dynamics of multiple interacting species (\cite{marrow_etal1996, law_etal1997,doebeli_dieckmann2000}), but the figure shows that such dynamics can also occur in the adaptive dynamics of a single species with multidimensional phenotype. The occurrence of cyclic evolutionary dynamics in a single species implies that the corresponding adaptive dynamics model is not equivalent to a symmetric competition model (even if it is an asymmetric competition model, as in the example above).

\vskip 1cm
\section{Local universality of symmetric competition models for determining evolutionary stability}
\vskip 0.5cm
So far we have considered the possible equivalence of general single-species adaptive dynamics models in a given dimension to the adaptive dynamics of symmetric competition models. If the adaptive dynamics convergence to a singular point, the question of evolutionary stability of the singular point arises, and we may then ask whether the evolutionary stability of singular points of generic single-species adaptive dynamics models can be understood in terms of the evolutionary stability of singular points in symmetric competition models. In this section, we show that in fact, the evolutionary stability of singular points of any generic single-species adaptive dynamics models can always be understood by means of symmetric competition models. Consider an arbitrary invasion fitness function $f(x,y)$ in the neighbourhood of a singular point $x^*$, which we assume to be $x^*=0$ without loss of generality. Because $f(x^*,x^*)=0$ as always, and 

\noindent $\partial f(x^*,y)/\partial y\vert_{y=x^*}=(\partial f(x^*,y)/\partial y_1\vert_{y=x^*},...,\partial f(x^*,y)/\partial y_m\vert_{y=x^*})=0$ by assumption of singularity of $x^*$, the Taylor expansion of the function $f$ in $x$ and $y$ in a neighbourhood of the point $(x^*,x^*)$ has the following form:
\begin{align}
f(x,y)=g(x)+xAy^T+\frac{1}{2}yHy^T+h.o.t.
\end{align}
where $g(x)$ is some function of $x$, $A$ is a square matrix, and $H$ is the symmetric Hessian matrix
\begin{align}
H=\begin{pmatrix} 
\left.\frac{\partial^2 f(x,y)}{\partial y_1^2}\right\vert_{y=x=x^*} & \ldots & \left.\frac{\partial^2 f(x,y)}{\partial y_1\partial y_m}\right\vert_{y=x=x^*} \\
 & \ldots & \\ 
 \left.\frac{\partial^2 f(x,y)}{\partial y_1\partial y_m}\right\vert_{y=x=x^*}  & \ldots & \left.\frac{\partial^2 f(x,y)}{\partial y_m^2}\right\vert_{y=x=x^*}  
 \end{pmatrix}
\end{align}
Here $h.o.t.$ denotes terms of order $>2$ in $y$, and $x^T$ and $y^T$ denote the transpose of the vectors $x=(x_1,...,x_m)$ and $y=(y_1,...,y_m)$, where $m$ is the dimension of phenotype space. 
In generic models the evolutionary stability of the singular point is determined by the Hessian matrix $H$ (\cite{leimar2009, doebeli2011}). Here we call a model generic if the Hessian $H$ is non-degenerate (i.e., has a trivial kernel).

Now consider the modified invasion fitness function
\begin{align}
\tilde f(x,y)=g(x)+\frac{1}{2}yHy^T,
\end{align}
also defined in a neighbourhood of $x^*$. Then the Jacobian $\tilde J$ of the corresponding selection gradient $\tilde s$ is simply the matrix $H$, which by construction is symmetric and hence has real eigenvalues. In particular, the linear dynamics defined by $\tilde s$ has no cyclic component, and hence $\tilde s$ is a gradient field (\cite{spivak1999}). It therefore follows from Proposition 2.1 that the invasion fitness function $\tilde f$ is equivalent to the invasion fitness of a symmetric competition model. On the other hand, the evolutionary stability of the singular point $x^*$ for the invasion fitness function $\tilde f$ is also given by the matrix $H$, and hence is exactly the same as the evolutionary stability of the singular point $x^*$ for the original invasion fitness function $f$. This proves the following 
\begin{proposition}
Given any single-species invasion fitness function $f$ with a non-degenerate Hessian matrix $H$, and a singular point $x^*$ of the corresponding adaptive dynamics model, then in a neighbourhood of $x^*$, there is an invasion fitness function $\tilde f$ derived from a symmetric competition model for which $x^*$ is a singular point, and such that the evolutionary stability of $x^*$ is the same for $f$ and $\tilde f$.\qed
\end{proposition}
It is important to note that while the evolutionary stability is the same for $f$ and $\tilde f$, convergence stability of $x^*$ in the adaptive dynamics models defined by $f$ and $\tilde f$ (i.e., by $s$ and $\tilde s$ in the construction above) is generally not the same. We note that even if it were, convergence stability generally depends on the mutational variance-covariance matrix (\cite{leimar2009}). Nevertheless, the proposition shows that the evolutionary stability of singular points of any single species adaptive dynamics model can be fully understood in terms of the evolutionary stability of singular points in symmetric competition models. We again note that evolutionary stability is {\it not} affected by the mutational variance-covariance matrix even in high-dimensional phenotype spaces (\cite{leimar2009}), hence this result is important for generalizing results already known for evolutionary stability in competition models, such as the finding in \cite{doebeli_ispolatov2010} that increasing the dimension of phenotype space generally increases the likelihood of evolutionary branching in symmetric competition models.

\vskip 1cm
{\bf \large Conclusions}
\vskip 0.5cm

We have shown that any model for the adaptive dynamics of a single species that is defined on a simply connected phenotype space and satisfies the gradient condition (19) is equivalent to an adaptive dynamics model for symmetric frequency-dependent competition. Specifically,  expression (15) can be considered a ``normal form'' for any given invasion fitness function satisfying the gradient condition (19). The obvious advantage of having such a normal form is that results obtained for the normal form are general and hold for any adaptive dynamics model that can be transformed into this normal form. To illustrate this, we have shown that any single-species adaptive dynamics model in a 1-dimensional phenotype space is equivalent to a symmetric competition model. In addition, we have shown that the evolutionary stability of single-species adaptive dynamics models with arbitrary phenotypic dimension can be understood in terms of symmetric competition models. Attempts at finding normal forms of invasion fitness functions have been made previously. For example, \cite{durinx_etal2008} showed that for every single-resident fitness function there exists a Lotka-Volterra competition model that has the same single-resident invasion fitness function. However, the interaction function was not partitioned into a frequency-dependent competition kernel and a frequency-independent carrying capacity. The approach presented here appears to be at the same time simpler and more specific, which is probably due to the fact that \cite{durinx_etal2008} dealt with the more complicated issue of normal forms describing the transition from monomorphic to polymorphic populations. Our normal form (15) is simpler because it only considers selection gradients in monomorphic populations, and hence only requires the definition of the generalized carrying capacity function (17) and the generalized competition kernel (18). Even though our normal form is only valid for invasion fitness functions satisfying (19) and for the monomorphic resident population, for those conditions it is general because it holds globally, i.e., everywhere in phenotype space, rather than just in the neighbourhood of singular points. And it is more specific because it disentangles the frequency-dependent and the frequency-independent components of selection. Generalized competition function without separation of frequency-dependent and frequency-independent components have also been considered by \cite{meszena_etal2005} for polymorphic populations. 

Essentially, the normal form (15) holds for any frequency-dependent adaptive dynamics model whose selection gradient is a gradient field, and hence whose dynamics can be described as a hill-climbing process on a fixed landscape. Because such hill-climbing processes cannot capture oscillatory behaviour, this makes it clear that the normal form cannot hold for any adaptive dynamics model exhibiting cyclic dynamics, e.g. evolutionary arms races in predator-prey systems (\cite{doebeli_dieckmann2000, doebeli2011}). To illustrate this, we have given an example of cyclic dynamics in a single-species adaptive dynamics model for asymmetric competition in 2-dimensional phenotype space, which therefore does not have the normal form (15). In general, the normal form can be applied to models of single species with high-dimensional phenotype spaces, and it is important to note that the generalized carrying capacity (17) and competition kernel (18) may be complicated functions in general. For example, the generalized carrying capacity may have multiple local maxima and minima (each representing a singular point of the adaptive dynamics), and the generalized competition kernel may have positive curvature at $x=y$ (representing positive frequency dependence). Accordingly, the adaptive dynamics resulting from a normal form may  exhibit repellers and dependence on initial conditions. 

The normal form (15) does not generally apply to the adaptive dynamics of multiple species. For example, if a single species adaptive dynamics model has a normal form (15), and if that normal form predicts evolutionary branching, then evolutionary branching does indeed occur in the given adaptive dynamics model, but the normal form cannot be used to derive the adaptive dynamics after evolutionary diversification has occurred. This is because to derive the adaptive dynamics ensuing after evolutionary branching, one has to know the ecological attractor of coexisting phenotypes, i.e., one has to have explicit information about the ecological dynamics underlying the adaptive dynamics model, which the normal form does not contain. We leave it as a challenge for future research to derive normal forms for the adaptive dynamics of multiple interacting species. 

\vskip 1cm
\noindent {\bf Acknowledgements:} We thank P. Krapivsky and M. Plyushchay for discussions. M.D. acknowledges the support of NSERC (Canada) and of the Human Frontier Science Program. I. I. acknowledges the support of FONDECYT (Chile).

\vskip 1cm
\noindent {\bf Author contributions:} M.D and I. I. contributed equally to this work.

\newpage
\bibliography{logistic}
\bibliographystyle{prslb}

\newpage
\begin{figure}[!]
\includegraphics[width=0.5\textwidth]{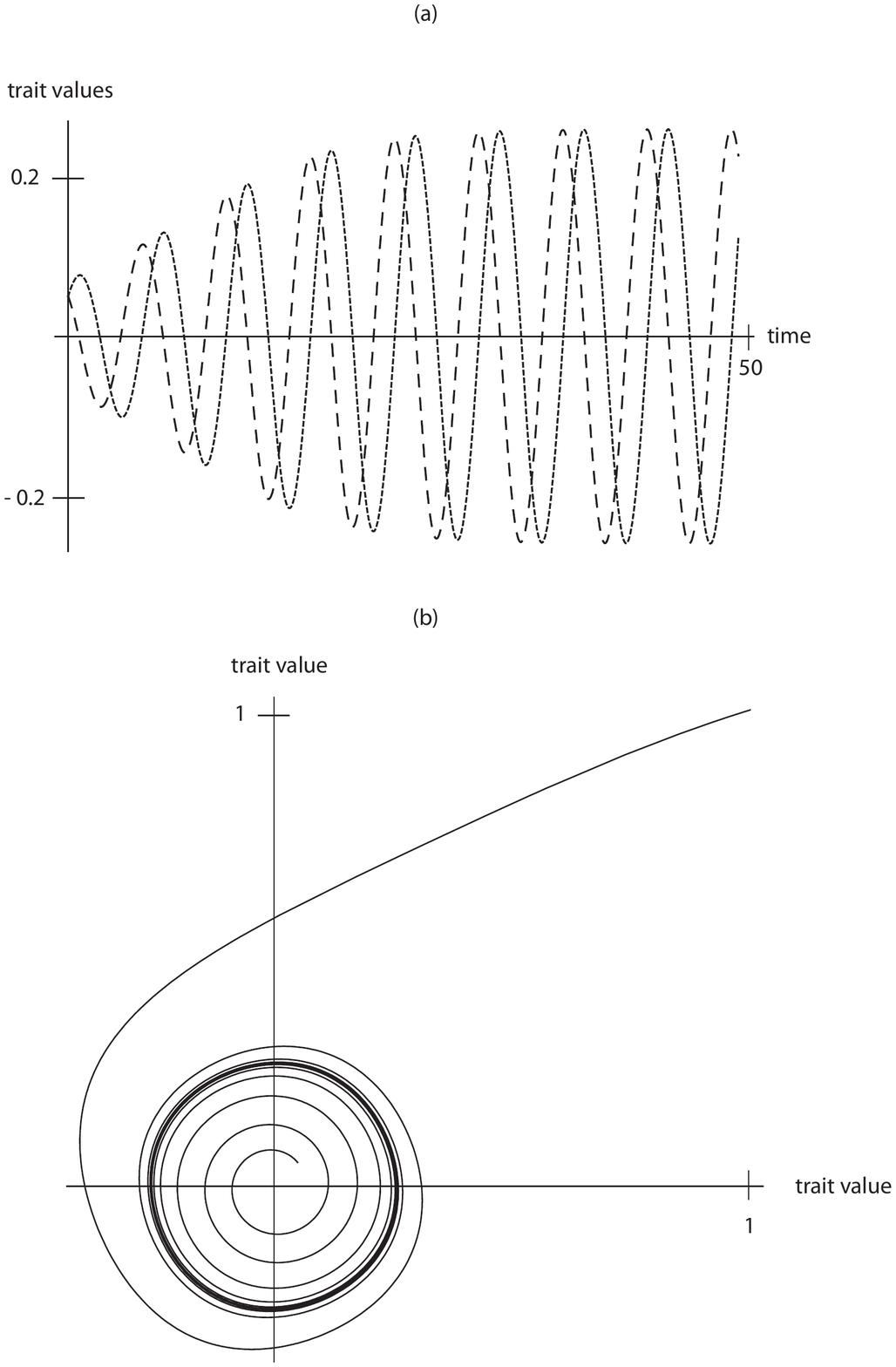}
\end{figure} 

\noindent {\bf \large Figure legend}

\vskip 1cm
\noindent{\bf Figure 1:} Example of cyclic adaptive dynamics in a single species with 2-dimensional phenotype space. The figure shows a numerical solution of the dynamical system 
$dx_1/dt=s_1(x_1,x_2)$ and $dx_2/dt=s_2(x_1,x_2)$, 
where $s_1$ and $s_2$ are given by (33) and (34). This system reflects the simplifying assumption that the mutational variance-covariance matrix (\cite{leimar2009, doebeli2011}) is the identity matrix. 
Panel (a) shows the two traits $x_1$ and $x_2$ as a function of time, and panel (b) shows the corresponding phase diagram, illustrating convergence to a limit cycle from two different initial conditions both inside and outside the limit cycle. Parameter values were $\sigma_K=\sigma_\alpha=1$, $c_1=-1$ and $c_2=-0.1$.


\end{document}